# The role of the coherence in the cross-correlation analysis of diffraction patterns from two-dimensional dense mono-disperse systems

T. Latychevskaia[1], G. F. Mancini[2] and F. Carbone[2]*

[1]Physics Department, University of Zurich, Winterthurerstrasse 190, 8057 Zurich, Switzerland.
[2]Laboratory for Ultrafast Microscopy and Electron Scattering, Institute of Condensed Matter Physics, École Polytechnique Fédérale de Lausanne, CH-1015 Lausanne, Switzerland.
*Corresponding author: Fabrizio Carbone, fabrizio.carbone@epfl.ch

**Abstract:** The investigation of the static and dynamic structural properties of colloidal systems relies on techniques capable of atomic resolution in real space and femtosecond resolution in time. Recently, the cross-correlation function (CCF) analysis of both X-rays and electron diffraction patterns from dilute and dense aggregates has demonstrated the ability to retrieve information on the sample's local order and symmetry. Open questions remain regarding the role of the beam coherence in the formation of the diffraction pattern and the properties of the CCF, especially in dense systems. Here, we simulate the diffraction patterns of dense two-dimensional monodisperse systems of different symmetries, varying the transverse coherence of the probing wave, and analyze their CCF. We study samples with different symmetries at different size scale, as for example, pentamers arranged into a four-fold lattice where each pentamer is surrounded by triangular lattices, both ordered and disordered. In such systems, different symmetry modulations are arising in the CCF at specific scattering vectors. We demonstrate that the amplitude of the CCF is a fingerprint of the degree of the ordering in the sample and that at partial transverse coherence, the CCF of a dense sample corresponds to that of an individual scattering object.

# Introduction

The X-ray or electron diffraction pattern of a single crystal exhibits distinct peaks whose intensities and positions can be used to deduce the crystal symmetry and structure. For such a crystallographic experiment, the transverse coherence of the probing wave needs to be just larger than the interatomic distances. This allows the coherent addition of the waves scattered from adjacent atoms which in turn, depending on the phase difference, results in either constructive or destructive interference in the far-field. The large number of atoms in a crystal ensures the high intensity of the peaks created by the constructive interference. On the other hand, the diffraction from a non-crystalline disordered system gives a diffraction pattern that exhibits concentric rings at characteristic scattering vectors which are attributed to the geometrical parameters of the unit cell. The recovery of the sample distribution or even its basic parameters, such as the symmetry, from the diffraction pattern is a more complex task than in the case of a single crystal.

In 2009, Wochner *et al.* demonstrated that the analysis of the angular cross-correlation functions in a diffraction pattern can reveal hidden symmetries in a non-crystalline sample[1]. They studied a colloidal suspension of polymethylmethacrylate (PMMA) spheres of 117 nm in radius by recording its diffraction pattern with partially coherent X-ray radiation of 0.154 nm wavelength. The diffraction pattern exhibited rings typical of a disordered system. However, the cross-correlation of the intensities along the azimuthal angle revealed pronounced modulations at certain scattering vectors that corresponded to characteristic periodicities in the sample. With their work Wochner *et al.* triggered the application of X-ray Cross-Correlation Analysis (XCCA) for revealing local symmetries in disordered systems[2-6].

The idea of using the angular cross-correlation functions (below we refer to it as cross-correlation function) for the determination of the structure of single particles in dilute solutions was originally proposed by Kam in 1977[7]. In Kam's approach, a spherical harmonics expansion of the scattered amplitudes was employed to recover the structure of an individual particle in solution. However, this approach was not fully explored until recently when it has been revised both theoretically[4-5,8-12] and experimentally[10,13-16]. It has been shown that the diffraction pattern of an individual particle can be extracted from the average over many diffraction patterns of a diluted ensemble of randomly oriented particles. This method uses the assumption that the interference between the waves scattered from different particles is negligible. To achieve this condition, thousands of coherent diffraction patterns with just a few particles in the scene (diluted sample) are summed up. Thus, the effect of the interference between the wave scattered from the particles is averaged out in the limit of an infinite number of images. The method of single-particle structure retrieval by the calculation of the cross-correlation functions is considered to be an alternative approach to a crystallographic experiment without the need of having a crystal.

In general, structure retrieval methods using the cross-correlation analysis of a diffraction pattern require partial transverse (spatial) coherence of the probing beam. In practice electron and X-ray sources have partial coherence. Here, and throughout in the text we refer to transverse coherence as coherence. The effect of infinite and partial coherence on the XCCA and single particle retrieval methods for two-dimensional *dilute* samples have been studied by Kurta *et. al* [3]. They have shown that the diffraction pattern of a sample consisting of 121 randomly distributed and randomly rotated pentagons (300 nm in size) exhibits no distinct peaks but only rings when it is acquired at infinite coherence. The effect of the partial coherence was studied on a sample consisting of 11 pentagons preferably oriented and therefore exhibiting peaks in the diffraction pattern. The coherence length was selected to be 1200 nm, 600 nm and 300 nm. As the coherence decreased, the contribution from the interference between the particles decreased, and as a result, peaks associated with the local structure got higher contrast. However, the effects of partial coherence have not been studied on dense samples of many particles, which is a practically interesting situation.

In our work we study dense and dilute samples of relatively large number of identical particles. We consider spherical particles of 5 nm in diameter arranged into domains and into an ordered lattice. We also consider the same spherical particles but assembled into pentamers that are randomly distributed and rotated thorough the dense sample. Pentamers are interesting five-fold symmetry object of investigation, which have been of particular interest since Wochner *et al.* reported odd symmetries in the CCF[1]. We provide a detailed numerical study of the diffraction patterns and of the related CCFs for dense two-dimensional systems at different coherence of the probing wave, from infinite coherence to the coherence length comparable to the size of single particle.

## Methods

### CCF definition, properties and amplitude

An example of a dense system is shown in Fig. 1 (a), which is an experimental image of an alkanethiol-capped gold nanoparticles supracrystal deposited on a copper grid covered with an amorphous carbon substrate, taken with Trasmission Electron Microscopy (TEM). Figure 1(b) exhibits a cartoon model of a selected fragment of the gold nanoparticles arrangement with their interdigitated ligands. In this disordered sample, both nanoparticles and ligand atoms can be arranged into sub-ordered structures, see for example Fig. 1(c). Though a Debye-Scherrer diffraction pattern is expected from such a disordered sample, the cross-correlation analysis can reveal the symmetries present in the sample, as for example, the periodical arrangement of the atoms. The cross-correlation analysis requires partial coherence of the probing beam. For this reason, we study the diffraction patterns of this supracrystal at the partial coherence of the probing wave $L_{coh}$ = 5, 10, 20 nm and infinity. The relative sizes of the examined coherence lengths are color coded in Fig. 1(d). An intensity distribution in the diffraction pattern at a certain scattering vector *s* is extracted and shown in Fig. 1(e). It is worth pointing out that already the intensity distribution contains a modulation specific to the sample symmetry, but this information is often buried under the noise and can only be revealed by calculating the CCF from the azimuthal intensity distribution, see Fig. 1(f).

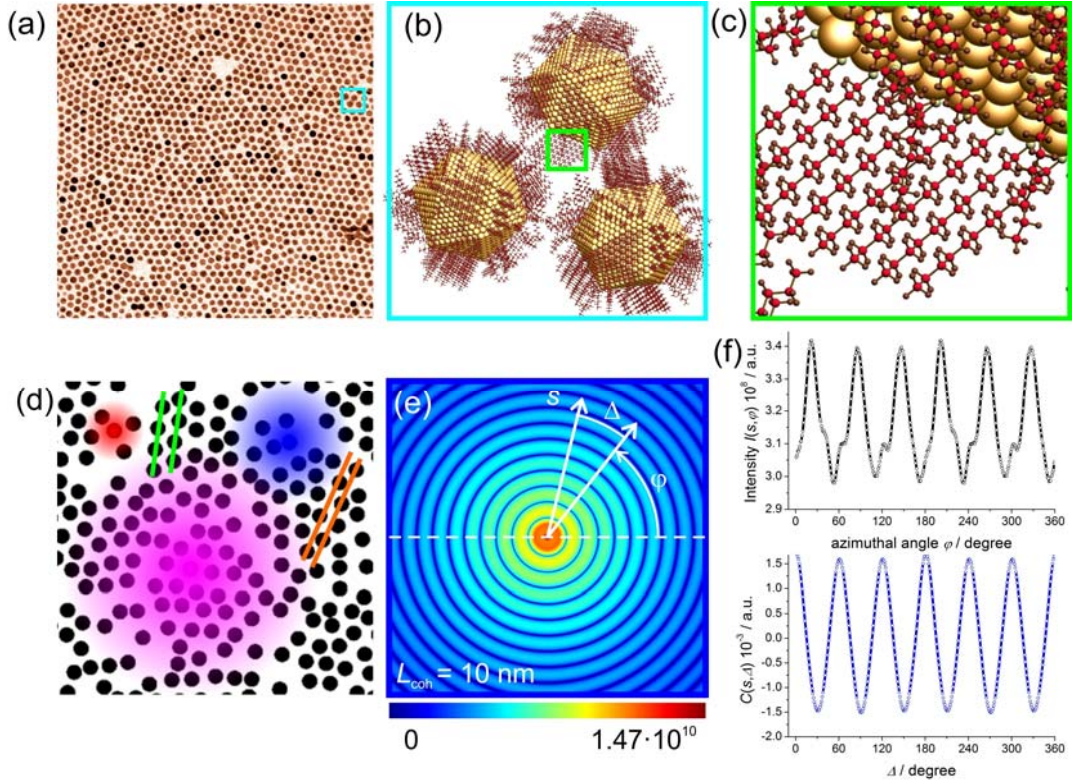

**Figure 1.** Illustration to the definition of the cross-correlation function. (a) TEM image of the functionalized gold nanoparticles supracrystal on an amorphous carbon substrate[17]. (b) A model of a selected fragment of the gold nanoparticles arrangement shown in the cyan square in panel (a). (c) Arrangement of the atoms in the ligands attached to the gold nanoparticle surface, obtained as a magnification of the green square in panel (b). (d) Simulated sample composed of spheres of 5 nm in diameter and arranged into randomly rotated domains of 40 nm size. The sample fragment is 100 nm × 100 nm in size and the coloured spots indicate coherence area: $L_{coh}$ = 20 nm (magenta), $L_{coh}$ = 10 nm (blue) and $L_{coh}$ = 5 nm (red). The green lines show the crystallographic planes separated by a distance $d_0$. The orange lines show the crystallographic planes separated by the distance $d_1$. (e) Illustration to the definitions of the symbols for the CCF calculation. (f) Azimuthal intensity distribution $I(s,\varphi)$ and its cross-correlation function $C(s,\Delta)$. Note the symmetry: $C(s,\Delta) = C(s,-\Delta)$ and $C(s,\pi + \Delta) = C(s,\pi - \Delta)$.

We define the CCF as proposed by Wochner *et al.*[1]:

$$C(s,\Delta) = \frac{\langle I(s,\varphi)I(s,\varphi+\Delta)\rangle_\varphi - \langle I(s,\varphi)\rangle_\varphi^2}{\langle I(s,\varphi)\rangle_\varphi^2} \qquad (1)$$

where $s$ is the component of the scattering vector $\vec{s}$ that is perpendicular to the direction of the incident beam, $\varphi$ is the azimuthal angle at a certain $s$, as illustrated in Fig. 1(e), and $\langle ... \rangle$ means averaging over $\varphi$:

$$\langle I(s,\varphi) \rangle_\varphi = \frac{1}{2\pi} \int_{-\pi}^{\pi} I(s,\varphi) \, \mathrm{d}\varphi. \qquad (2)$$

Some of the properties of $C(s,\Delta)$ include:

(1) $C(s,\Delta)$ has always maximum at $C(s,0)$.

(2) $C(s,\Delta)$ is always symmetrical at $\Delta = 0$ and $\Delta = \pi$:

$$\begin{aligned} C(s,\Delta) &= C(s,-\Delta) \\ C(s,\pi+\Delta) &= C(s,\pi-\Delta). \end{aligned} \qquad (3)$$

These equations hold also for an experimental intensity function $I(s,\varphi)$ contaminated by noise as illustrated in Fig. 1(f).

(3) If the function $I(s,\varphi)$ is centro-symmetric, meaning that $I(s,\varphi) = I(s,\pi+\varphi)$, its $C(s,\Delta)$ always has maximum at $\pi$: $C(s,\pi) = C(s,0)$.

(4) If the function $I(s,\varphi)$ is not centro-symmetric so that $I(s,\varphi) \neq I(s,\pi+\varphi)$, its $C(s,\Delta)$ does not have maximum at $\pi$, and can have minimum at $\pi$.

The amplitude of the CCF can be a critical parameter when the CCF is extracted from noisy experimental data. The total intensity in the far-field can be represented as following:

$$I_{total} = |U_1 + U_2 + ... U_P|^2 \qquad (4)$$

where $U_p$ is the complex-valued scattered wave by particle $p$, where $p = 1..P$. The total number of particles is $P$. Equation 4 can be re-written in expanded form:

$$\begin{aligned} I_{total} &= |U_1|^2 + |U_2|^2 + ... |U_P|^2 + \\ &+ U_1^* U_2 + U_2^* U_1 + U_1^* U_3 + U_3^* U_1 ... = \\ &= \sum_{p=1}^{P} |U_p|^2 + \sum_{\substack{i,j=1, \\ i \neq j}}^{P} U_i^* U_j + U_j^* U_i = \sum_{p=1}^{P} |U_p|^2 + \text{Interference term} = \sum_{p=1}^{P} I_p(s,\varphi) + \text{Interference term} \end{aligned}$$

$$(5)$$

where the intensity distributions are re-written in $(s, \varphi)$ coordinates. In Eq. 5, the second term is due to the interference between the waves scattered off the different particles. For very short coherence length, in an extreme case- comparable to the size of the particle, the second sum "Interference term" can be neglected and only the first sum remains. We use this approximation ("Interference term" = 0) in the following formulas.

Next, we assume that the particles are identical but randomly rotated. The intensity of the wave scattered from a single particle is given by $I_0(s,\varphi)$ and the intensity of that from a rotated particle is given by $I_0(s,\varphi+\alpha)$. The total intensity is then:

$$I_{total}(s,\varphi) = \sum_{p=1}^{P} I_p(s,\varphi) = \sum_{p=1}^{P} I_0(s,\varphi + \alpha_p). \tag{6}$$

The Fourier expansion of the diffraction pattern of single particle is given by:

$$I_0(s,\varphi) = \sum_m f_m(s)\, e^{-im\varphi}. \tag{7}$$

By substituting Eq.7 into Eq.6 we obtain[16]:

$$I_{total}(s,\varphi) = \sum_{p=1}^{P} I_0(s,\varphi + \alpha_p) = \sum_{p=1}^{P} \sum_m f_m(s)\, e^{-im\varphi} e^{-im\alpha_p}. \tag{8}$$

To calculate the CCF of the particles ensemble, given by Eq.1, we write its components:

$$\begin{aligned}
\langle I_{total}(s,\varphi)\rangle_\varphi &= \frac{1}{2\pi}\int_{-\pi}^{\pi} I_{total}(s,\varphi)\,d\varphi = \frac{1}{2\pi}\int_{-\pi}^{\pi}\left(\sum_{p=1}^{P} I_0(s,\varphi+\alpha_p)\right)d\varphi = \\
&= \frac{1}{2\pi}\sum_{p=1}^{P}\int_{-\pi}^{\pi} I_0(s,\varphi+\alpha_p)\,d\varphi = \frac{1}{2\pi}P\int_{-\pi}^{\pi} I_0(s,\varphi)\,d\varphi = P\langle I_0(s,\varphi)\rangle_\varphi,
\end{aligned} \tag{9}$$

$$\begin{aligned}
\langle I_{total}&(s,\varphi)I_{total}(s,\varphi+\Delta)\rangle_\varphi = \frac{1}{2\pi}\int_{-\pi}^{\pi} I_{total}(s,\varphi)I_{total}(s,\varphi+\Delta)\,d\varphi = \\
&= \frac{1}{2\pi}\int_{-\pi}^{\pi}\left(\sum_{p=1}^{P}\sum_m f_m(s)\,e^{-im\varphi}e^{-im\alpha_p}\right)\left(\sum_{p'=1}^{P}\sum_k f_k(s)\,e^{-ik\varphi}e^{-ik\Delta}e^{-ik\alpha_{p'}}\right)d\varphi = \\
&= \left(\sum_{p=1}^{P}\sum_m f_m(s)\,e^{-im\alpha_p}\right)\left(\sum_{p'=1}^{P}\sum_k f_k(s)\,e^{ik\Delta}e^{ik\alpha_{p'}}\right)\delta(k+m) = \sum_{p,p'=1}^{P}\sum_m f_m(s)f_{-m}(s)\,e^{-im(\alpha_p-\alpha_{p'})}e^{im\Delta},
\end{aligned}$$
(10)

where

$$\sum_{p,p'=1}^{P} e^{-im(\alpha_p - \alpha_{p'})} \approx P, \tag{11}$$

and

$$\langle I_0(s,\varphi)I_0(s,\varphi+\Delta)\rangle_\varphi = \frac{1}{2\pi}\int_{-\pi}^{\pi} I_0(s,\varphi)I_0(s,\varphi+\Delta)\mathrm{d}\varphi =$$
$$= \frac{1}{2\pi}\int_{-\pi}^{\pi}\left(\sum_m f_m(s)\mathrm{e}^{-im\varphi}\right)\left(\sum_k f_k(s)\mathrm{e}^{-ik\varphi}\mathrm{e}^{-ik\Delta}\right)\mathrm{d}\varphi = \quad (12)$$
$$= \sum_m f_m(s)\sum_k f_k(s)\mathrm{e}^{im\Delta}\delta(k+m) = \sum_m f_m(s)f_{-m}(s)\mathrm{e}^{im\Delta}$$

Thus, the result of Eq. 10 can be written as:

$$\langle I_{\text{total}}(s,\varphi)I_{\text{total}}(s,\varphi+\Delta)\rangle_\varphi = P\langle I_0(s,\varphi)I_0(s,\varphi+\Delta)\rangle_\varphi. \quad (13)$$

We substitute the results of Eq. 9 and Eq. 13 into Eq. 1 and obtain the CCF:

$$C(s,\Delta) = \frac{P\langle I_0(s,\varphi)I_0(s,\varphi+\Delta)\rangle_\varphi - \left(P\langle I_0(s,\varphi)\rangle_\varphi\right)^2}{\left(P\langle I_0(s,\varphi)\rangle_\varphi\right)^2} \sim \frac{\langle I_0(s,\varphi)I_0(s,\varphi+\Delta)\rangle_\varphi}{P}. \quad (14)$$

Thus, according to Eq. 14, the CCF of the particles ensemble is proportional to the CCF of the single particle and inversely proportional to the number of particles $P$. This, however, is only true when the coherence effects can be neglected or when the coherence length is comparable to the particle size, and in a dense sample that approximation holds if the coherence length is shorter than the particle-particle distance. When the sample is dense and the probing beam is partially coherent, higher values of the CCF amplitude are expected.

**Simulations at partial coherence**

The coherence properties of the probing beam are conventionally characterized by the mutual coherence function, or the complex coherence factor:

$$\mu(\vec{r}_1,\vec{r}_2) = \frac{J(\vec{r}_1,\vec{r}_2)}{\sqrt{I(\vec{r}_1)I(\vec{r}_2)}}. \quad (15)$$

In Eq. 15, $J(\vec{r}_1,\vec{r}_2)$ is the mutual intensity function $J(\vec{r}_1,\vec{r}_2) = \langle E(\vec{r}_1,t)E(\vec{r}_2,t)\rangle_T$, where the averaging is performed over a time $T$ which is much longer than the fluctuation time of $E(\vec{r}_1,t)$, $I(\vec{r}_1)$ and $I(\vec{r}_2)$ are the intensity values of the incoming beam at points $\vec{r}_1$ and $\vec{r}_2$.

The diffraction pattern obtained with a partial coherence can be directly simulated as the convolution[3,18-19]:

$$I(\vec{s}) = I_{coh}(\vec{s}) \otimes \mu(\vec{s}), \tag{16}$$

where $I_{coh}(\vec{s})$ is the diffraction pattern obtained at infinite coherence, and $\mu(\vec{s})$ is the Fourier transform of the mutual coherence. Though the exact mutual coherence function is often complicated, it can be assumed in the form of a Gaussian distribution[20]:

$$\mu(\vec{r}_1, \vec{r}_2) = \exp\left[-(\vec{r}_1 - \vec{r}_2)^2 / 2L_{coh}^2\right], \tag{17}$$

where $L_{coh}$ is the coherence length. Using Eqs. 16 and 17, Vartanyants et al.[18] have simulated the partial coherent diffraction pattern of a diluted sample consisting of particles arranged into a two-dimensional array. The diffraction pattern obtained at an infinite coherence length exhibited Bragg peaks, and at a coherence length comparable to the size of the single particle, the diffraction pattern turned into the diffraction pattern of a single particle. Those simulations thus relate well to the typical experimental observations. However, in the case of a dense sample, Eqs. 16 and 17 fail to correctly simulate the diffraction pattern at partial coherence, as illustrated below. Therefore, for the simulations of the diffraction patterns at limited coherence length, we avoid the convolution and apply the particle-by-particle algorithm explained below.

**Diffraction pattern of a monodisperse sample**

The diffraction pattern of the entire sample is given by:

$$I(\vec{s}) = \left| \sum_{p=1}^{P} o_p(\vec{r}) \exp(-i\vec{s}\vec{r}) \right|^2, \tag{18}$$

where $o_p(\vec{r})$ is the distribution of the particle $p$ and $P$ is the total number of particles. Assuming that all particles are identical, from which $o_p(\vec{r}) = o(\vec{r})$, we can re-write Eq. 18:

$$I(\vec{s}) = \left| \sum_{p=1}^{P} o(\vec{r} - \vec{r}_p) \exp(-i\vec{s}\vec{r}) \right|^2, \tag{19}$$

where $\vec{r}_p$ are the coordinates of the particle $p$. By substitution $\vec{r} - \vec{r}_p = \vec{r}'$, we obtain:

$$I(\vec{s}) = \left| \sum_{p=1}^{P} o(\vec{r}') \exp(-i\vec{s}\vec{r}') \exp(-i\vec{s}\vec{r}_p) \right|^2 = \left| o(\vec{r}') \exp(-i\vec{s}\vec{r}') \right|^2 \left| \sum_{p=1}^{P} \exp(-i\vec{s}\vec{r}_p) \right|^2, \tag{20}$$

which simplifies to

$$I(\vec{s}) = \left| o(\vec{r}) \exp(-i\vec{s}\vec{r}) \right|^2 \left| \sum_{p=1}^{P} \exp(-i\vec{s}\vec{r}_p) \right|^2. \tag{21}$$

Equation 21 describes the diffraction pattern of a sample consisting of identical particles. The first factor in Eq. 21 is the diffraction pattern of an individual particle. The second factor in Eq. 21 in the case of a coherence length $L_{coh}$ comparable to the size of one particle is just the number of particles $P$. In the case of an infinite coherence, the diffraction pattern can be calculated simply by Fourier transform of the sample distribution.

**Simulation of diffraction patterns at partial coherence**

In the case of a dense sample, Eqs. 16 and 17 fail to correctly simulate the diffraction pattern at partial coherence, as illustrated in Fig. 2. The sample is shown in Fig. 2(a) and (b), and its diffraction pattern at infinite coherence length of the probing wave is shown in Fig. 2(c). When the coherence length is equal to the size of the particle, the expected diffraction pattern is the diffraction pattern of a single particle multiplied by the number of particles in the sample, as follows from Eq. 21. Thus, for the sample consisting of 5 nm spheres, when $L_{coh}$ = 5 nm, the diffraction pattern should exhibit concentric rings. This situation is correctly simulated by the particle-by-particle algorithm (explained below), as shown in Fig. 2(d) and it is not correctly simulated by applying the convolution given by Eq. 16 and 17, as shown in Fig. 2(f) and (h). At $L_{coh}$ = 5 nm, the distinct peaks are still observed in the simulated diffraction pattern, see Fig. 2(f). At $L_{coh}$ = 2.5 nm, the peaks are blurred out, however no concentric rings but some six-fold structure is emerging, see Fig. 2(h). To validate the correctness of the convolution routine given by Eq. 16 and 17, we simulated diffraction pattern of the same sample as in the work by Vartanyants and Robinson[18] at infinite and partial coherence and verified that the diffraction pattern indeed turned into the diffraction pattern of a single particle when the coherence is decreased. Thus we concluded that the method of simulating the diffraction patterns at partial coherence by convolution gives correct results only in the case of diluted samples. Therefore, for the dense samples investigated in this work we applied the particle-by-particle algorithm as explained below.

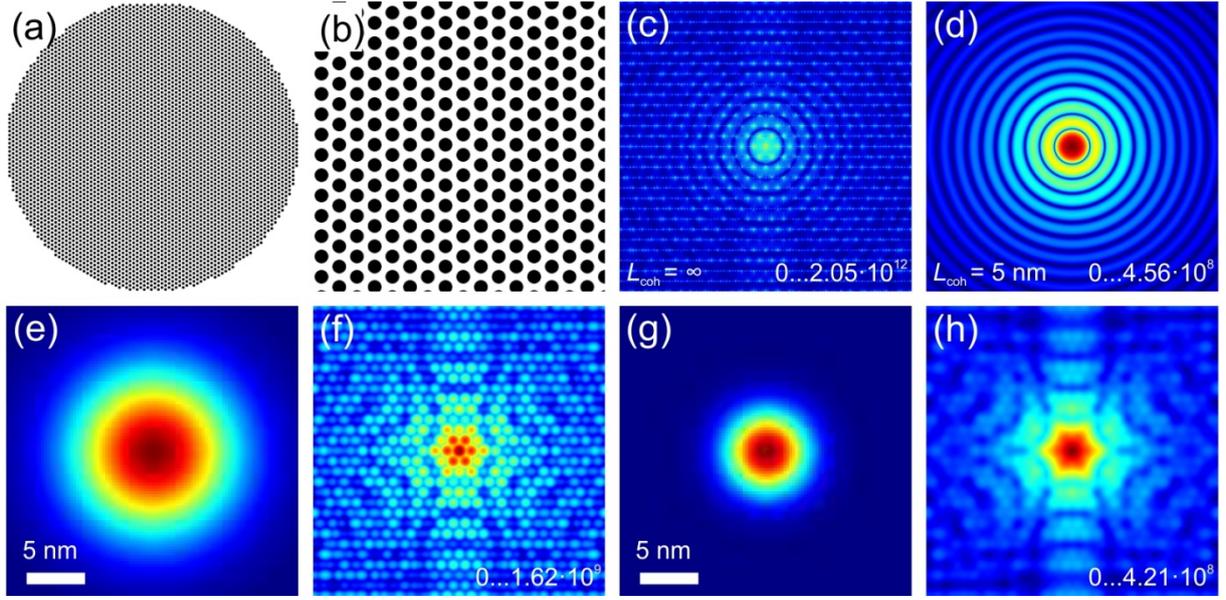

**Figure 2.** Simulated diffraction pattern of spheres with 5 nm diameter with a partial coherence of $L_{coh}$ = 5 nm. (a) Entire sample (500 nm × 500 nm in size) used for calculations. (b) Magnified sample fragment of 100 nm × 100 nm in size. (c) Diffraction pattern of the sample simulated at infinite coherence. (d) Diffraction pattern of the sample simulated at partial coherence $L_{coh}$ = 5 nm by the particle-by-particle algorithm. (e) Mutual coherence simulated with Eq. 17 with $L_{coh}$ = 5 nm; an area of 25 nm × 25 nm is shown. (f) Diffraction pattern simulated with Eq. 16 using the mutual coherence with $L_{coh}$ = 5 nm. (g) Mutual coherence simulated with Eq. 17 at $L_{coh}$=2.5 nm, the area of 25 nm × 25 nm is shown. (h) Diffraction pattern simulated with Eq. 16 using mutual coherence with $L_{coh}$ = 2.5 nm.

**Particle-by-particle simulation of diffraction patterns in the case of partial coherence**

Each particle $p$ is assigned its $(x_p, y_p)$ coordinates (in pixels) and thus, two one-dimensional arrays containing $x$ and $y$ coordinates are created: $x_p$ and $y_p$. The sum given by Eq. 21 is calculated over all the particles as following.

(i) For the first particle, $p =1$, the self-interference term is calculated:

$$|U_1|^2 = |\exp(-i\vec{s}\vec{r}_1)|^2 = 1. \tag{22}$$

(ii) The coordinates of the first particle $(x_1, y_1)$ are compared with the coordinates of every other particle in the array, $p = 2, 3...$ :

$$l = \sqrt{(x_1 - x_p)^2 + (y_1 - y_p)^2} \leq 4L_{coh}, \quad p = 2,3...P. \tag{23}$$

Only the particles whose coordinates are within $4L_{coh}$ distance from the particle $p$ are considered. $4L_{coh}$ is selected as the cut-off distance, as according to the mutual coherence distribution given by Eq. 17, at this distance the degree of coherence is $\mu = \exp\left[-(4L_{coh})^2 / 2L_{coh}^2\right] = 3.35 \cdot 10^{-4}$ only and the input of the interference term is negligible. For those particles, whose coordinates satisfy Eq. 23, the interference term is calculated:

$$\exp\left[-(\vec{r}_p - \vec{r}_1)^2 / 2L_{coh}^2\right] \cdot \left(U_p^* U_1 + U_p U_1^*\right), \tag{24}$$

where $(\vec{r}_p - \vec{r}_1)^2 = (x_p - x_1)^2 + (y_p - y_1)^2$ and $U_p^* U_1 + U_p U_1^* = 2\cos\left\{\frac{2\pi}{N}\left[(x_p - x_1)^2 + (y_p - y_1)^2\right]\right\}$ and $N$ is the number of pixels. The results obtained in (i) and (ii) are added.

(iii) The coordinates of the particle $p = 1$ are eliminated from the array $x_p$ and $y_p$, and thus the previously second particle becomes the first particle. The loop returns to step (i).

## Results

**Sample consisting of spheres of 5 nm in diameter**

a. **Spheres arranged into randomly rotated domains**

The sample simulated here is mimicking the two-dimensional supracrystal of gold nanoparticles of about 5.7 nm in diameter separated by 7.63 nm studied experimentally by Mancini et al.[17]. This simulated dense sample consists of $P = 13404$ spheres of 5 nm in diameter arranged into perfect triangular lattice with sphere-to-sphere distance of $d = 7$ nm. The sample size is 1000 nm × 1000 nm and its distribution is sampled with 4000 × 4000 pixels. Within the sample, round domains of 40 nm in diameter are created (326 domains in total); the position of each domain is randomly generated. Within each domain, the particle distribution is rotated and the rotation angle is also randomly generated. In addition, each particle is randomly shifted from its position by up to 1 nm. The distribution of a portion of the sample is shown in Fig. 1(d), which also gives an idea about the compatibility between the particles, domain sizes and the coherence length.

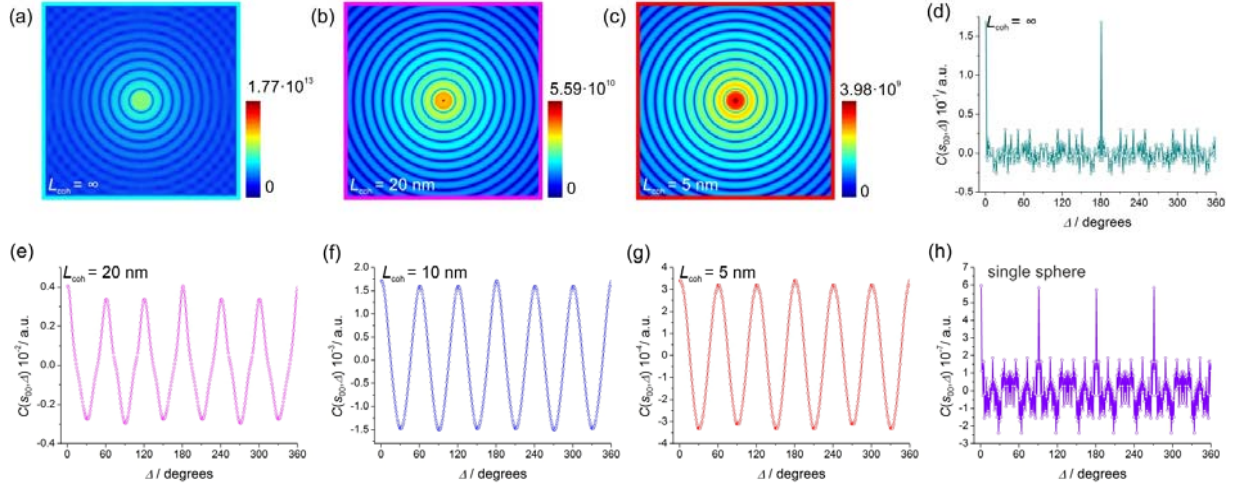

**Figure 3.** Simulated diffraction patterns of spheres of 5 nm diameter arranged into randomly rotated 40 nm domains at (a) infinite coherence and at partial coherence (b) $L_{coh}$ = 20 nm and (c) $L_{coh}$ = 5 nm. (d) – (g) Related distributions of CCF at $s_{00}=2\pi/d_0$=1.03 nm$^{-1}$, at the coherence length $L_{coh}$ = ∞, 20 nm, 10 nm and 5 nm. (h) CCF calculated from the diffraction pattern of an individual sphere.

The diffraction patterns of the sample simulated at different coherence length are shown in Fig. 3(a) – (c). At any coherence length, no distinct peaks are observed. In the case of a coherence length larger than the sample area dimension, which we call an infinite coherence length $L_{coh}$ = ∞, the diffraction pattern is simulated simply by the Fourier transform of the transmission function of the sample (electron density of the sample in the case of X-ray imaging), as shown in Fig. 3(a). At infinite coherence length, no modulations are observed in the CCF, as shown in Fig. 3(d). The two spikes at $\Delta$ = 0° and $\Delta$ = 180° are numerical artefacts as the azimuthal intensity distribution is matching itself when shifted by 0° and 180°. At partial coherence, six-fold modulations are clearly observed in the CCF at low $s$, see Fig. 3(e) – (g). These modulations are caused by the arrangement of the spheres. Spheres within one domain are arranged as triangular lattice, which contains crystallographic planes that are apart from each other by $d_0 = 7 \cdot \sqrt{3}/2$ nm $= 6.06$ nm and by $d_1 = 7/2$ nm $= 3.5$ nm; the crystallographic planes are shown in Fig. 1(d). This gives the related $s$ values $s_{00} = 2\pi/d_0 = 1.03$ nm$^{-1}$ and $s_{01} = 2\pi/d_1 = 1.8$ nm, at which the modulations are observed. Even at low coherence of $L_{coh}$ = 5 nm, where the diffraction pattern resembles that of a single particle, the characteristic six-fold modulations are still observed in CCF, see Fig. 3(c) and (g). This is because a probing wave with $L_{coh}$ = 5 nm still provides sufficient coherence to cause interference effects between neighbouring particles.

The amplitude of the CCF, as it can be seen from Fig. 3(h), is about $2 \cdot 10^{-7}$ a. u. According to the theoretical prediction for the amplitude of the CCF given by Eq. 14, the expected maximum is given by the CCF of the single particle/$P$ and equals to $2 \cdot 10^{-7}/13404 = 1.5 \cdot 10^{-11}$. However, the CCF of a single particle, shown in Fig. 3(h) exhibits a very low amplitude. This is because the diffraction pattern of a perfect sphere consists of perfect rings without any modulations and as a consequence there are no modulations in the CCF. The signal in the CCF of a single particle is just numerical noise due to the finite sampling. However, we can assume that the amplitude of the CCF of a single particle, shown in Fig. 3(h), is around 1 and with this approximation the expected amplitude of the CCF, according to Eq. 14, is about $1/P = 7.53 \cdot 10^{-5}$. Thus, the theory predicts the CCF values to be in the range: $7.53 \cdot 10^{-5} - 1.5 \cdot 10^{-11}$. However, from the graphs we see the following amplitudes of the CCF:

$L_{coh}$ = 20 nm: amplitude of CCF $4.0 \cdot 10^{-3}$ a.u.,

$L_{coh}$ = 10 nm: amplitude of CCF $\approx 1.7 \cdot 10^{-3}$ a.u., and

$L_{coh}$ = 5 nm: amplitude of CCF $\approx 3.5 \cdot 10^{-4}$ a.u.

These values are orders of magnitude higher than expected. However, they can be explained if we consider each domain of ordered particles as a "particle" itself. This, according to Eq.14, gives for the amplitude of the CCF approximately 1/number of domains = 1/326 = 0.003, which agrees better with the observed amplitudes of the CCFs. It is interesting to note that the scattering vector $s_{00}$, where the modulations are observed, is related to the distance between the crystallographic planes *within* a domain or, in other words, to the distances between the particles. But the amplitude of the CCF is proportional to 1/number of domains.

Thus, our explanation for the large CCF values is the presence of local ordering throughout the entire sample. To verify this idea we provide another example of a completely ordered sample.

### b. Spheres arranged into a perfect triangular lattice

The sample here consists of $P = 4617$ spheres arranged into a perfect triangular lattice with a sphere-to-sphere distance of $d = 7$ nm. The CCF plotted at the same $s_{00} = 1.03$ nm$^{-1}$ are shown in Fig. 4.

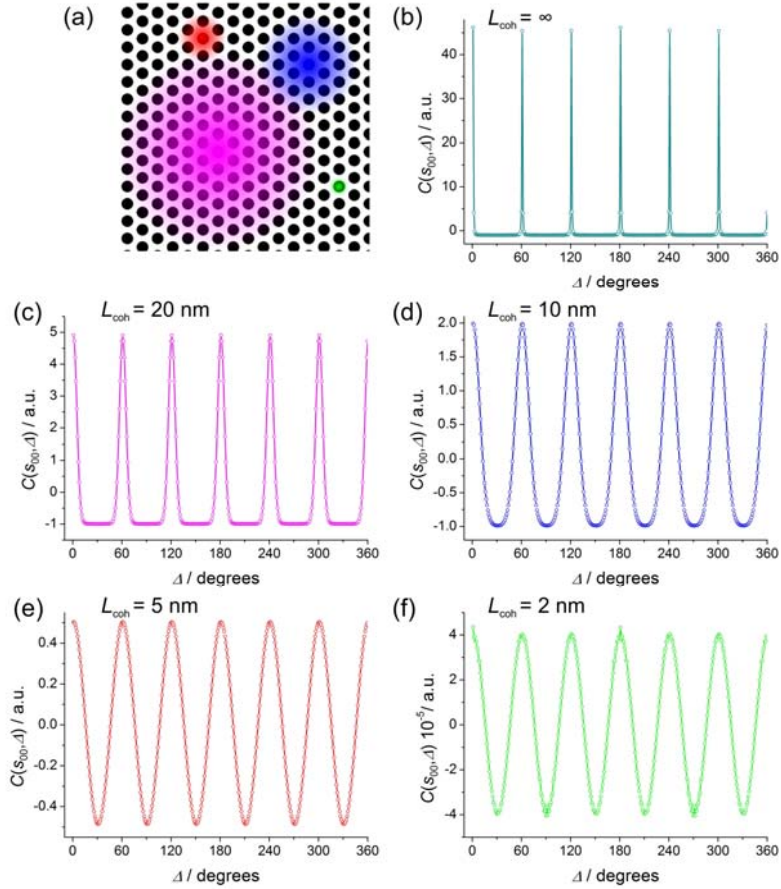

**Figure 4.** Dense sample of spheres arranged into a perfect triangular lattice. (a) Sample fragment of 100 nm × 100 nm in size, the coloured spots indicate the coherence area: $L_{coh}$ = 20 nm (magenta), $L_{coh}$ = 10 nm (blue), $L_{coh}$ = 5 nm (red) and $L_{coh}$ = 5 nm (green). (b) – (f) CCFs obtained from the diffraction patterns of the spheres of 5 nm diameter arranged into a perfect triangular lattice at $s_{00} = 2\pi/d_0 = 1.03$ nm$^{-1}$, at the coherence length (b) $L_{coh} = \infty$, (c) 20 nm, (d) 10 nm, (e) 5 nm and (f) 2 nm.

The expected amplitude of the CCF, according to Eq. 14, amounts to approximately $2 \cdot 10^{-7}/4617 = 4.3 \cdot 10^{-11}$. And again, if the amplitude of the CCF calculated for a single sphere was around 1, the expected amplitude of the CCF of the sample would be $1/P = 2.2 \cdot 10^{-4}$. Thus, the amplitude of the CCF should be in the range: $2.2 \cdot 10^{-4} - 4.3 \cdot 10^{-11}$. However, from the graphs plotted in Fig. 4 we see the following amplitudes:

$L_{coh}$ = 20 nm: amplitude of CCF ≈ 5,

$L_{coh}$ = 10 nm: amplitude of CCF ≈ 2,

$L_{coh}$ = 5 nm: amplitude of CCF ≈ 0.5, and

$L_{coh}$ = 2 nm: amplitude of CCF ≈ $4 \cdot 10^{-5}$.

These are very high values, which cannot be explained by Eq.14. Equation 14 can only explain the result at $L_{coh}$ = 2 nm, where the interference effects are truly negligible, the very approximation employed in Eq.14.

Thus, the amplitude of the CCF can have higher values than expected from the rule $1/P$, and the reasons are:

(1) (amplitude of CCF ~ $1/P$) only holds when the interference effect can be neglected. This is not the case when even weak coherence is present.

(2) Presence of ordered regions can also dramatically affect CCF amplitude. In this case, the amplitude of the CCF can be proportional to 1/number of ordered regions.

These observations can explain the relatively high values of the amplitude of CCF for dense samples reported previously[1,17]. Wochner *et al.* reported CCFs which are showing modulations in the range of -0.02 to +0.02. This would mean that the number of particles in the probing volume is about $1/P$ = 1/0.02 ~ 50 particles. In their experiment, Wochner *et al.* had probing beam of 10 microns in diameter and 100 nm in diameter spheres arranged into 12 spheres clusters (each cluster is approximately 250 nm in diameter) distributed in three-dimensional volume[1]. Thus, these experimental parameters provide much larger number of particles in the probing volume than just 50 particles. This mismatch can be explained by our observations that the simple rule CCF ~ $1/P$ does not apply in the case of a dense system probed with a partially coherent beam.

**Pentamers**

Next, we consider the same spherical particles of 5 nm in diameter assembled into pentamers, as shown in Fig. 4(a). Five-fold symmetry is an interesting object of investigation, as it is forbidden in the long-range in crystals but can be found on the short-range scales within disordered systems[21]. Five-fold symmetries have also been of particular interest since Wochner *et al.* reported odd symmetries in the CCF[1], which Wochner *et al.* explained by the fact that the Ewald sphere at high scattering vectors *s* cannot be approximated by a plane, and therefore deviations from the Friedel's law are possible. However, the observed by Wochner *et al.* five-fold symmetry is found at relatively low *s*. Moreover, the symmetries of the CCF at adjacent *s* values are even-fold. This would imply that the Friedel's law only fails at some very specific scattering vectors. The remarkable observation of odd symmetry modulations in the CCF triggered further research on two-dimensional distributions of five-fold symmetrical objects[2-6], reporting only even symmetries in CCFs extracted from their diffraction patterns. Altarelli *et al.* showed that for two-dimensional disordered systems only even Fourier components of the CCF give nonzero contributions[2]. The diffraction pattern of a three-dimensional distribution of odd-symmetrical objects (oxygen clusters) was simulated by Kurta *et al.*[5], showing that odd symmetries can be observed when the curvature of the Ewald sphere is taken into account. In our simulations the pentamers are arranged with different distributions: (i) randomly distributed and rotated pentamers (ii) ordered pentamers surrounded by a fine periodic lattice and (iii) randomly distributed and rotated pentamers surrounded by a fine periodic lattice.

### a. Pentamers of spheres of 5 nm, randomly distributed

In this subsection we consider a sample consisting of $P = 843$ pentamers. The pentamers are randomly distributed and randomly rotated throughout the sample of 500 nm × 500 nm in size, see Fig. 5(a). The CCFs are calculated from the simulated diffraction patterns simulated at partial coherence $L_{coh} = $ 20 nm, 10 nm and 5 nm are shown in Fig. 5 (b) – (d).

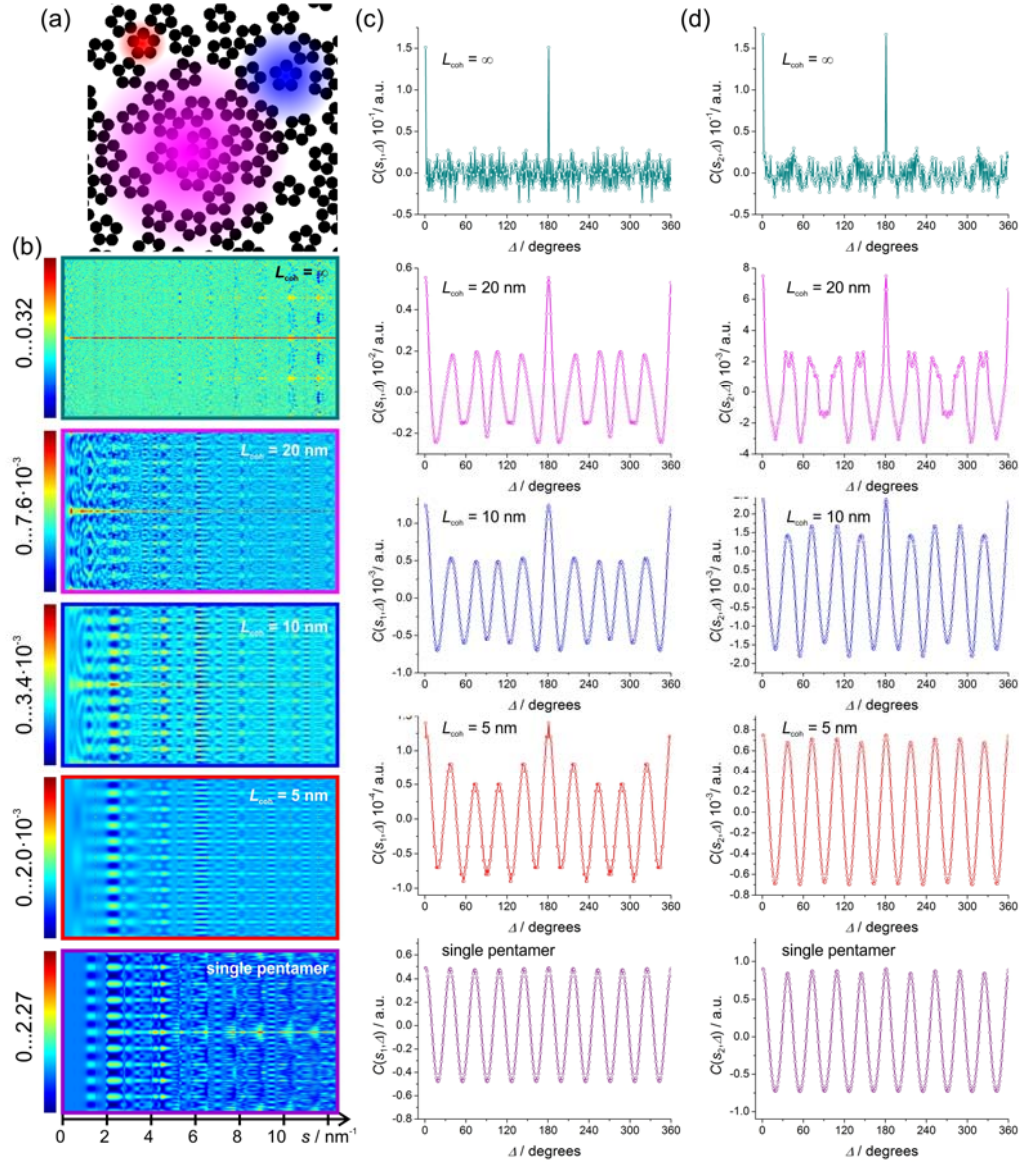

**Figure 5.** Simulated diffraction patterns of randomly distributed and rotated pentamers and related CCFs. (a) Sample fragment of 100 nm × 100 nm in size, the coloured spots indicate coherence area: $L_{coh} = 20$ nm (magenta), $L_{coh} = 10$ nm (blue) and $L_{coh} = 5$ nm (red). (b) Two-dimensional plots of the CCF as functions of $s$ vectors (abscissa) and $\Delta$ (ordinate) calculated from the diffraction patterns simulated at partial coherence $L_{coh} = 20$ nm, $L_{coh} = 10$ nm

and $L_{coh}$ = 5 nm. (c) One-dimensional plots of the CCF at $s_1$ = 1.4 nm$^{-1}$ and (d) $s_2$ = 2.4 nm$^{-1}$ for related $L_{coh}$.

No distinct peaks are observed in any of the simulated diffraction patterns. At infinite coherence length of the probing wave, also no modulations are observed in the CCF distribution, see Fig. 5(b) – (d). However, at partial coherence, the CCFs exhibit ten-fold modulations pronounced at low $s$ values, as for example the maxima at $s_1$ = 1.4 nm$^{-1}$ and $s_2$ = 2.4 nm$^{-1}$ thus indicating the presence of five-fold symmetry in the sample, see Fig. 5(b) –(d). The scattering vectors $s_1$ = 1.4 nm$^{-1}$ and $s_2$ = 2.4 nm$^{-1}$ are related to the real-space length of 4.8 nm and 2.9 nm, respectively, which however do not match any length related to pentamers arrangement. However, the maxima of the intensity observed in the diffraction pattern at $s_1$ = 1.4 nm$^{-1}$ and $s_2$ = 2.4 nm$^{-1}$ relate well with the maxima of the intensity in the diffraction pattern from an aperture whose diameter equals to the outer diameter of the pentamer, in our case, 12 nm.

From the graphs plotted in Fig. 5 we see the following amplitudes at $s_1$:
$L_{coh}$ = 20 nm: amplitude of CCF ≈ 6·10$^{-3}$,
$L_{coh}$ = 10 nm: amplitude of CCF ≈ 1·10$^{-3}$, and
$L_{coh}$ = 5 nm: amplitude of CCF ≈ 1.5·10$^{-4}$,
which agrees well with 1/number of pentamers = 1/843 = 1.12·10$^{-3}$.

These results are in agreement with the observation by Kurta et al.[3]: as the coherence decreased, the peaks associated with the local structure become more pronounced. Though the pentamers are all randomly oriented and no peaks are observed in the diffraction pattern at characteristic $s$, the CCF analysis helps to extract modulations at characteristic $s$ which are related to the symmetry (five-fold) of the local structure. It is remarkable that at partial coherence, the CCF of the entire sample resembles the CCF of an individual pentamer, compare Fig. 5(b) – (d), which has been previously discussed by Vartanyants and Robinson[18]. Therefore, it should be possible to extract a diffraction pattern of an individual pentamer from the diffraction pattern of a dense sample via XCCA analysis, as previously mentioned by Kurta et al.[4].

**b. Pentamers of spheres of 5 nm surrounded by regular lattices, ordered**

To study the appearance of different symmetries under different coherence length conditions, we created a diluted sample organized as following. The pentamers ($P$ = 206) are organized into a 4-fold lattice with the distance between pentamers being 30 nm. The total sample size is 500 nm × 500 nm. Around each pentamer there is a triangular lattice of point scatterers, with the distance between two closest scatterers being 2 nm. The sample is shown in Fig. 6(a). The diffraction pattern at infinite coherence was simulated by the Fourier transform of the sample, the diffraction patterns at partial coherence were simulated by convolution with the mutual coherence function, as given by Eqs. 16 – 17. At infinite coherence length, see Fig. 6(b), the peaks from all symmetries are observed in the CCF: the six-fold peaks at large $s$, ten-fold peaks at intermediate $s$ and four-fold peaks at small $s$. As the coherence length decreases, the four-fold peaks disappear, but the six- and ten-fold peaks remain, see Fig. 6(b) – (d). The ten-fold modulations are observed at the same $s$ values at which they were observed in the case of the pentamers sample: $s_1 = 1.4$ nm$^{-1}$ and $s_2 = 2.4$ nm$^{-1}$, see Fig. 6(c). The fine lattice around each pentamer is a perfect triangular lattice which contains crystallographic planes that are apart from each other by $d_0 = 2 \cdot \sqrt{3}/2$ nm $= 1.73$ nm, and $d_1 = 2/2$ nm $= 1.0$ nm. The related $s$ values are thus: $s_3 = 2\pi/d_0 = 3.63$ nm$^{-1}$ and $s_4 = 2\pi/d_1 = 6.28$ nm. Also, the second order of diffraction from the planes $d_0$ is observed at $s_5 = 7.26$ nm$^{-1}$. At these $s$ values, six-fold modulations are pronounced even at low coherence $L_{coh} = 5$ nm, see Fig. 6(d).

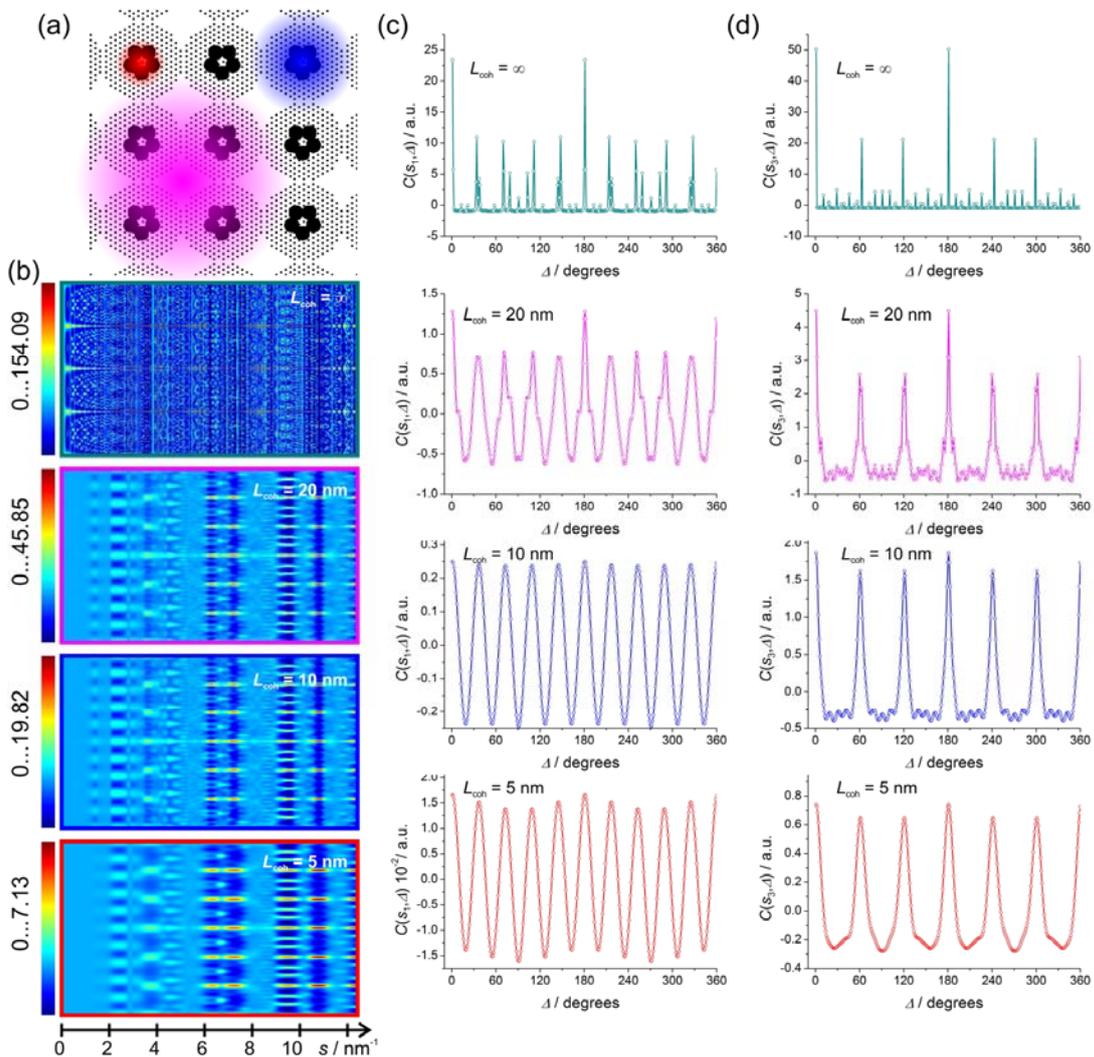

**Figure 6.** Simulated diffraction patterns of ordered pentamers surrounded by regular, ordered lattices and their CCFs. (a) Magnified sample fragment of 100 nm × 100 nm in size, the coloured spots indicate coherence area: $L_{coh}$ = 20 nm (magenta), $L_{coh}$ = 10 nm (blue) and $L_{coh}$ = 5 nm (red). (b) Two-dimensional plots of the CCF as functions of $s$ vectors (abscissa) and $\Delta$ (ordinate) calculated from the diffraction patterns simulated at partial coherence $L_{coh}$ = 20 nm, $L_{coh}$ = 10 nm and $L_{coh}$ = 5 nm. (c) One-dimensional plots of the CCF at $s_1$ = 1.4 nm$^{-1}$ and (d) $s_3$ = 3.63 nm$^{-1}$ for related $L_{coh}$.

### c. Pentamers of spheres of 5 nm surrounded by regular lattices, disordered

The results of studying a sample with random positions and rotation of the pentamers distributed within a fine lattice structure ($P$ = 133) are shown in Fig. 7. The sample here is mimicking the situation of atoms ordered around nanoparticles, as for example atoms ordered in ligands attached to the surface of gold nanoparticles[17].

As the coherence decreases, the modulations caused by the symmetry in the local structure emerge, see Fig. 7(b) – (d). The ten-fold modulations related to the five-fold symmetry of the pentamers are found at the same $s$ as in the case of ordered pentamers: $s_1$ = 1.4 nm$^{-1}$ and $s_2$ = 2.4 nm$^{-1}$, see Fig. 7(c). The twelve-fold modulations related to the six-fold symmetry of the fine lattices are found at the same $s$ as in the previous example of ordered lattices: $s_3$ = $2\pi/d_0$ = 3.63 nm$^{-1}$, $s_4$ = $2\pi/d_1$ = 6.28 nm and $s_5$ = 7.26 nm$^{-1}$, see Fig. 7(d).

It is therefore possible to detect the symmetry of the local structure by means of CCF analysis of the diffraction pattern, provided that the diffraction pattern is recorded at partial coherence. Despite the fact that sub-structures are randomly distributed and randomly rotated throughout the sample, their intrinsic ordering produces well-pronounced modulations in the CCF.

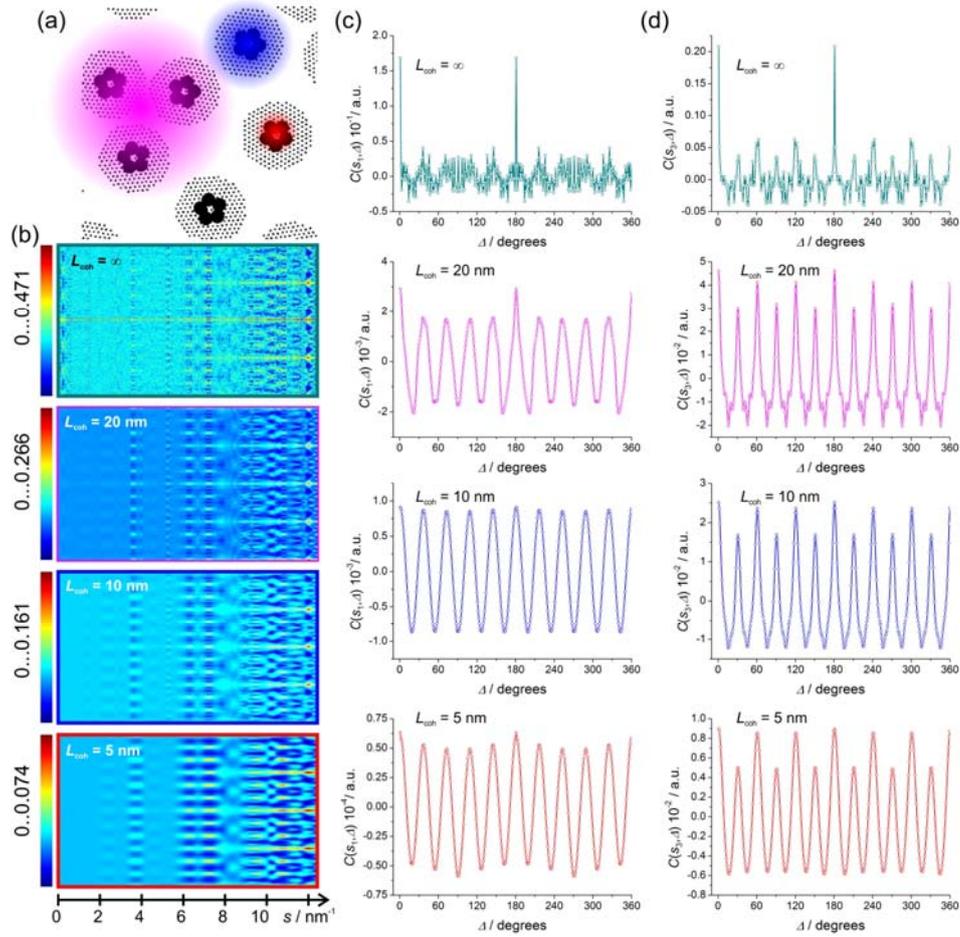

**Figure 7.** Simulated diffraction patterns of disordered pentamers surrounded by regular, disordered lattices and their CCFs. (a) Magnified sample fragment of 100 nm × 100 nm in size, the coloured spots indicate coherence area: $L_{coh}$ = 20 nm (magenta), $L_{coh}$ = 10 nm (blue) and $L_{coh}$ = 5 nm (red). (b) Two-dimensional plots of the CCF as functions of *s* vectors (abscissa) and $\Delta$ (ordinate) calculated from the diffraction patterns simulated at partial coherence $L_{coh}$ = 20 nm, $L_{coh}$ = 10 nm and $L_{coh}$ = 5 nm. (c) One-dimensional plots of the CCF at $s_1$ = 1.4 nm$^{-1}$ and (d) $s_3$ = 3.63 nm$^{-1}$ for the related $L_{coh}$.

## Discussion

We studied the diffraction patterns and their CCFs in dense samples under different coherence lengths of the probing wave. For dense samples we developed particle-by-particle algorithm for the simulation of the diffraction patterns, which allows obtaining more accurate results than those obtained by conventional approach by convolution with the mutual coherence function. For the studied dense samples, no peaks were observed in the diffraction patterns at any coherence length. However, at partial coherence, the characteristic symmetries of the sample were revealed by the cross-correlation analysis. We showed that as the coherence decreases, the modulation related to the ordering of local structure becomes more pronounced in the CCF. The sample with particles organized into domains (ordered subsystems), where each domain was randomly rotated, exhibited modulations in the CCF related to the ordering of the particles within a domain. We showed that the simple rule of the amplitude of the CCF~1/number of particles does not apply in the case of a dense system probed with partially coherent beam. In the case of such systems, the CCF has a much higher amplitude than predicted by the rule CCF~1/number of particles, and thus CCF can be easily measured for systems with a large number of particles. The modulations in the CCF are observed at the scattering vectors related to the distances between the periodically arranged particles; however the intensity of those modulations is inversely proportional to the number of ordered subsystems. Thus, the amplitude of the CCF is a measure of the degree of ordering in the system.

The dense sample of randomly distributed and rotated pentamers exhibited ten-fold modulations in the CCF at characteristic $s$. Moreover, the CCF distribution at partial coherence resembles that of a single pentamer, which indicates the possibility to extract the diffraction pattern of a single pentamer from a single-shot diffraction pattern of a dense sample.

To conclude, our study confirms that the cross-correlation analysis can be applied to study the arrangement of sub-systems in a disordered sample, revealing for example the ordering of the atoms in ligands attached to nanoparticles in two-dimensional supracrystals, even though the nanoparticles themselves are not arranged in a perfect lattice. In a dense sample, unlike in the case of diluted ones, even at a very low coherence length comparable to the size of particle, the interference effects between particles are not negligible. At infinite coherence of the probing beam these effects lead to a complete smearing out of the diffraction peaks. We demonstrated that at low coherence these effects lead to the appearance of peaks in the CCF that are a signature of certain symmetries in the sample.

## Acknowledgements

The work was funded by the Swiss National Science Foundation (SNSF) through the grant No. PP00P2–128269/1) and by the University of Zurich. The authors acknowledge J. Reguera for the experimental TEM image of the alkanethiol-capped gold nanoparticles supracrystal and F. Stellacci for useful discussions.